\documentclass[11pt,preprintnumbers,titlepage,nofootinbib]{revtex4-2}%
\usepackage{textcomp}
\usepackage[latin9]{inputenc}
\usepackage{color}
\usepackage{amsmath}
\usepackage{amssymb}
\usepackage[unicode=true,  bookmarks=false,  breaklinks=false,pdfborder={0 0
1},backref=section,colorlinks=false]{hyperref}
\usepackage{textcomp}
\usepackage{wasysym}
\usepackage{array}
\usepackage{multirow}
\usepackage{wasysym}
\usepackage{wasysym}
\usepackage{amsfonts}
\usepackage{subfigure}
\usepackage{cleveref}
\usepackage{listings}
\usepackage{xcolor}
\usepackage{array}
\usepackage{url}
\usepackage{color}
\usepackage{subfigure}
\usepackage{graphicx}%
\hypersetup{
colorlinks,linkcolor=blue,citecolor=blue,urlcolor=blue}
\makeatletter
\DeclareFontEncoding{LGR}{}{}

\ProvideTextCommand{\~}{LGR}[1]{\char126#1}
\DeclareTextSymbolDefault{\textquotedbl}{T1}

\@ifundefined{textcolor}{}{
\definecolor{BLACK}{gray}{0}
\definecolor{WHITE}{gray}{1}
\definecolor{RED}{rgb}{1,0,0}
\definecolor{GREEN}{rgb}{0,1,0}
\definecolor{BLUE}{rgb}{0,0,1}
\definecolor{CYAN}{cmyk}{1,0,0,0}
\definecolor{MAGENTA}{cmyk}{0,1,0,0}
\definecolor{YELLOW}{cmyk}{0,0,1,0}
\definecolor{ballblue}{rgb}{0.13, 0.67, 0.8}
\definecolor{bleudefrance}{rgb}{0.19, 0.55, 0.91}
\definecolor{blue(ncs)}{rgb}{0.0, 0.53, 0.74}
\definecolor{darkpastelgreen}{rgb}{0.01, 0.75, 0.24}
\definecolor{darkspringgreen}{rgb}{0.09, 0.45, 0.27}
\definecolor{denim}{rgb}{0.08, 0.38, 0.74}
\definecolor{electricviolet}{rgb}{0.56, 0.0, 1.0}
}

\makeatother
\begin{document}
\preprint{CTP-SCU/2024011}
\title{Polarized Image of a Synchrotron-emitting Ring in Einstein-Maxwell-scalar
Theory }
\author{Yiqian Chen$^{a,b}$}
\email{chenyiqian@ucas.ac.cn}
\author{Lang Cheng$^{a}$}
\email{chenglang@stu.scu.edu.cn}
\author{Peng Wang$^{a}$}
\email{pengw@scu.edu.cn}
\author{Haitang Yang$^{a}$}
\email{hyanga@scu.edu.cn}
\affiliation{$^{a}$Center for Theoretical Physics, College of Physics, Sichuan University,
Chengdu, 610064, China}
\affiliation{$^{b}$School of Fundamental Physics and Mathematical Sciences, Hangzhou
Institute for Advanced Study, University of Chinese Academy of Sciences,
Hangzhou, 310024, China}

\begin{abstract}
This study investigates polarized images of an equatorial synchrotron-emitting
ring surrounding hairy black holes within the Einstein-Maxwell-scalar theory.
Our analysis demonstrates qualitative similarities between the polarization
patterns of hairy black holes and Schwarzschild black holes. However, due to
the non-minimal coupling between the scalar and electromagnetic fields, an
increase in black hole charge and coupling constant can substantially amplify
polarization intensity and induce deviations in the electric vector position
angle. These effects may offer observational signatures to distinguish hairy
black holes from Schwarzschild black holes.

\end{abstract}
\maketitle
\tableofcontents

{}

{}

\section{Introduction}

The groundbreaking images of the supermassive black holes at the centers of
M87 and the Milky Way galaxies, obtained by the Event Horizon Telescope (EHT)
Collaboration
\cite{Akiyama:2019cqa,Akiyama:2019brx,Akiyama:2019sww,Akiyama:2019bqs,Akiyama:2019fyp,Akiyama:2019eap,Akiyama:2021qum,Akiyama:2021tfw,EventHorizonTelescope:2022xnr,EventHorizonTelescope:2022vjs,EventHorizonTelescope:2022wok,EventHorizonTelescope:2022exc,EventHorizonTelescope:2022urf,EventHorizonTelescope:2022xqj}%
, have ushered in a new era for testing general relativity in the strong-field
regime. These observations, featuring a central brightness depression
surrounded by an asymmetric bright ring, are attributed to synchrotron
emission originating from relativistic plasma near the black hole. To identify
black hole parameters from the data, a comprehensive library of ray-traced
General Relativistic Magnetohydrodynamic (GRMHD) images of simulated accretion
flows has been constructed. While some models were eliminated by the EHT
results, a significant number of GRMHD models were found to be consistent with
the total intensity data \cite{Akiyama:2019fyp}.

In addition to total intensity data, the $2017$ EHT observations also captured
polarimetric information, which is influenced by the emitting plasma, magnetic
field and spacetime curvature. This polarimetric data provides a unique
opportunity to gain deeper insights into the properties of black holes and
their surrounding environments. Over the past few decades, simulated polarized
images have been used to explore various phenomena, including accretion states
\cite{Shcherbakov:2010ki,Palumbo:2020flt}, the jet \cite{Moscibrodzka:2015pda}%
, Faraday rotation \cite{Moscibrodzka:2017gdx,Jimenez-Rosales:2018mpc} and
more
\cite{Schnittman:2009im,Dexter:2016cdk,Gold:2016hld,Moscibrodzka:2019adb,GRAVITY:2020gka,GRAVITY:2020hwn,Moscibrodzka:2021fxv,Zhang:2022klr}%
. However, the majority of these studies have relied on GRMHD, which, while
powerful, is computationally intensive and can limit comprehensive parameter
surveys. Furthermore, the GRMHD approach primarily focuses on accretion flows,
potentially constraining our ability to investigate relativistic effects
around black holes.

Recently, a simplified model of a synchrotron-emitting fluid ring orbiting a
Schwarzschild black hole was proposed to better understand polarized images of
black holes \cite{EventHorizonTelescope:2021btj}. This model, with
appropriately chosen parameters, effectively reproduces key features of the
polarimetric morphology observed in EHT images of M87$^{\ast}$, such as the
Electric Vector Position Angle (EVPA) pattern and the relative polarized
intensity. Additionally, it can predict the polarization evolution caused by
orbital hotspots during flares, similar to those observed from Sgr A$^{\ast}$
with GRAVITY \cite{GRAVITY:2018ofz}. This model was later extended to Kerr
black holes \cite{Gelles:2021kti}, leading to further investigations of
polarized images using this and similar models within various theoretical
frameworks
\cite{Qin:2021xvx,Delijski:2022jjj,Qin:2022kaf,Lee:2022rtg,Zhu:2022amy,Liu:2022ruc,Hu:2022sej,Qin:2023nog,Deliyski:2023gik}%
.

Meanwhile, a class of Einstein-Maxwell-scalar (EMS) models, featuring a
non-minimal coupling between the scalar and Maxwell fields, has been
introduced to investigate the formation of hairy (scalarized) black holes
\cite{Herdeiro:2018wub}. Through fully nonlinear numerical simulations, these
EMS models have revealed the intriguing phenomenon of spontaneous
scalarization in Reissner-Nordstr\"{o}m (RN) black holes, driven by a
tachyonic instability. This process has spurred extensive research into the
properties and implications of scalarized RN black holes
\cite{Myung:2018jvi,Myung:2018vug,Brihaye:2019gla,Brihaye:2019dck,Fernandes:2019kmh,Fernandes:2019rez,Myung:2019oua,Peng:2019cmm,Zou:2019bpt,Astefanesei:2020qxk,Blazquez-Salcedo:2020jee,Blazquez-Salcedo:2020nhs,Fernandes:2020gay,Guo:2020zqm,Mai:2020sac,Myung:2020dqt,Myung:2020ctt,Myung:2020etf,Zou:2020zxq,Guo:2021zed,Wang:2020ohb,Zhang:2021etr,Zhang:2021nnn,Zhang:2022cmu,Jiang:2023yyn,Chen:2023eru}%
. In particular, recent studies have explored the optical appearance of
accretion disks surrounding scalarized RN black holes, demonstrating multiple
bright rings of varying radii and a noticeable increase in the total intensity
flux in black hole images \cite{Gan:2021pwu,Gan:2021xdl}. However, polarized
images of scalarized RN black holes remain unexplored. Investigating such
images could deepen our understanding of how interactions between scalar and
electromagnetic fields influence light propagation, potentially offering a
promising avenue for the search of hairy black holes.

This paper investigates polarized images of a synchrotron-emitting ring around
scalarized RN black holes, as well as those of RN black holes. The subsequent
sections of this paper are structured as follows: Sec. \ref{sec:Set-Up}
briefly reviews scalarized RN black holes, discusses geometric optics and
introduces the equatorial emission model in the EMS model. Sec.
\ref{sec:Numerical-Result} presents the numerical results of polarized images
for both equatorial and vertical magnetic fields. Finally, Sec.
\ref{sec:Conclusions} summarizes our findings. We adopt the convention
$G=c=4\pi\epsilon_{0}=1$ throughout the paper.

\section{Set Up}

\label{sec:Set-Up}

This section begins by providing a concise review of the scalarized RN black
hole solution within the 4-dimensional EMS model. It then examines light
propagation within this framework using geometric optics approximations.
Subsequently, we introduce an equatorial emission ring model to analyze
polarized images of the black holes.

\subsection{Black Hole Solution}

The EMS model, as detailed in \cite{Herdeiro:2018wub}, incorporates a gravity
theory with a scalar field $\phi$ and the electromagnetic field $A_{\mu}$
through the action,
\begin{equation}
S=\frac{1}{16\pi}\int d^{4}x\sqrt{-g}\left[  \mathcal{R}-2\partial_{\mu}%
\phi\partial^{\mu}\phi-f\left(  \phi\right)  F_{\mu\nu}F^{\mu\nu}\right]  ,
\label{eq:Action}%
\end{equation}
where $\mathcal{R}$ is the Ricci scalar and $F_{\mu\nu}=\partial_{\mu}A_{\nu
}-\partial_{\nu}A_{\mu}$ is the electromagnetic field strength tensor. In the
EMS model, the scalar field $\phi$ is non-minimally coupled to the
electromagnetic field $A_{\mu}$ via the coupling function $f\left(
\phi\right)  $. For scalar-free black holes, such as RN black holes, to exist,
the coupling function must satisfy the condition $\left.  df\left(
\phi\right)  /d\phi\right\vert _{\phi=0}=0$
\cite{Herdeiro:2018wub,Fernandes:2019rez}. This study specifically examines
the exponential coupling function $f\left(  \phi\right)  =e^{\alpha\phi^{2}}$
with $\alpha>0$. In the RN black hole background, the equation of motion
governing the scalar perturbation $\delta\phi$ is given by
\begin{equation}
\left(  \square-\mu_{\text{eff}}^{2}\right)  \delta\phi=0,
\label{eq:delta phi}%
\end{equation}
where $\mu_{\text{eff}}^{2}=-\alpha Q^{2}/r^{4}$. Notably, a tachyonic
instability arises when the effective mass squared $\mu_{\text{eff}}^{2}$
becomes negative. This instability can induce spontaneous scalarization in the
scalar-free solution, resulting in the formation of hairy (scalarized) black
hole solutions \cite{Herdeiro:2018wub}.

To obtain scalarized RN black hole solutions, we adopt the following ansatz
for the metric, electromagnetic field and scalar field,
\begin{align}
ds^{2}  &  =-N\left(  r\right)  e^{-2\delta\left(  r\right)  }dt^{2}+\frac
{1}{N\left(  r\right)  }dr^{2}+r^{2}\left(  d\theta^{2}+\sin^{2}\theta
d\varphi^{2}\right)  ,\nonumber\\
A_{\mu}dx^{\mu}  &  =\Phi\left(  r\right)  dt\text{ and}\ \phi=\phi\left(
r\right)  .
\end{align}
Appropriate boundary conditions are imposed at the event horizon $r_{h}$ and
at spatial infinity as follows,
\begin{align}
N\left(  r_{h}\right)   &  =0\text{, }\delta\left(  r_{h}\right)  =\delta
_{0}\text{, }\phi\left(  r_{h}\right)  =\phi_{0}\text{, }\Phi\left(
r_{h}\right)  =0\text{,}\nonumber\\
N\left(  \infty\right)   &  =1\text{, }\delta\left(  \infty\right)  =0\text{,
}\phi\left(  \infty\right)  =0\text{, }\Phi\left(  \infty\right)
=\Psi\text{.} \label{eq:infinity condition}%
\end{align}
Here, $\delta_{0}$ and $\phi_{0}$ characterize the black hole solutions, while
$\Psi$ represents the electrostatic potential. By specifying $\delta_{0}$ and
$\phi_{0}$, we obtain scalarized RN black hole solutions with a non-trivial
scalar field $\phi$ using the shooting method implemented in the $NDSolve$
function of $Wolfram\text{ }Mathematica$. The black hole mass $M$ and charge
$Q$ are determined from the asymptotic behavior of the metric functions at
infinity,
\begin{align}
N\left(  r\right)   &  =1-\frac{2M}{r}+\cdots,\nonumber\\
\Phi\left(  r\right)   &  =\Psi-\frac{Q}{r}+\cdots.
\end{align}
For simplicity, all physical quantities are expressed in units of the black
hole mass by setting $M=1$ throughout the paper.

Intriguingly, scalarized RN black holes have been shown to possess two photon
spheres outside the event horizon within specific black hole parameter regimes
\cite{Gan:2021pwu}. This distinctive feature gives rise to unique phenomena,
including black hole images with intricate structures
\cite{Gan:2021xdl,Guo:2022muy,Chen:2022qrw,Chen:2023qic,Chen:2024ilc} and echo
signals \cite{Guo:2021enm,Guo:2022umh}. Furthermore, studies on superradiant
instabilities and the nonlinear stability of these double photon sphere black
holes have been conducted \cite{Guo:2023ivz,Guo:2024cts}. For a comprehensive
exploration of black holes with multiple photon spheres, we direct readers to
\cite{Guo:2022ghl}.

\subsection{Propagation of Light}

The non-minimal coupling between the scalar and electromagnetic fields can
influence light propagation in scalarized RN black holes. Following
\cite{Schwarz:2020jjh}, we employ the geometric optics approximation to derive
the equations governing light propagation. In the Lorentz gauge, $\nabla_{\mu
}A^{\mu}=0$, the equation of motion for the electromagnetic field becomes
\begin{equation}
\nabla^{\mu}f\left(  \phi\right)  \left(  \nabla_{\mu}A_{\nu}-\nabla_{\nu
}A_{\mu}\right)  +f\left(  \phi\right)  \left(  \square A_{\nu}-R_{\nu}%
^{\ \mu}A_{\mu}\right)  =0, \label{eq:maxEq}%
\end{equation}
where $R_{\nu}^{\ \mu}$ is the Ricci tensor. The geometric optics
approximation is valid when the electromagnetic wavelength is much smaller
than any other relevant scale, including the characteristic scales of the
background metric and scalar field. This permits the use of the ansatz
\begin{equation}
A^{\mu}\left(  x\right)  =\operatorname{Re}\left[  \bar{A}^{\mu}\left(
x\right)  e^{i\theta\left(  x\right)  /\epsilon}\right]  ,
\label{eq:waveAnsatz}%
\end{equation}
where $\bar{A}^{\mu}$ is the slowly evolving complex amplitude vector, and
$\theta$ is the rapidly oscillating phase. Here, $\epsilon$ is a small
parameter introduced to track various orders of terms, with the
geometric-optics limit corresponding to $\epsilon\ll1$. The $4$-wavevector
$k_{\mu}$ is then defined as
\begin{equation}
k_{\mu}\equiv\frac{1}{\epsilon}\partial_{\mu}\theta,
\end{equation}
representing the local direction of wave propagation.

Substituting the ansatz $\left(  \ref{eq:waveAnsatz}\right)  $ into Eq.
$\left(  \ref{eq:maxEq}\right)  $, the term of order $\epsilon^{-2}$ yields
the dispersion relation $k_{\mu}k^{\mu}=0$. This dispersion relation, together
with the property $\nabla_{\mu}k_{\nu}=\nabla_{\nu}k_{\mu}$, leads to
\begin{equation}
k^{\mu}\nabla_{\mu}k^{\nu}=0,
\end{equation}
indicating that the propagation of light rays is governed by null geodesics.
Furthermore, the term of order $\epsilon^{-1}$ gives
\begin{equation}
\nabla_{\mu}f\left(  \phi\right)  \left(  k^{\mu}\bar{A}_{\nu}-\bar{A}^{\mu
}k_{\nu}\right)  +f\left(  \phi\right)  \left(  2k^{\mu}\nabla_{\mu}\bar
{A}_{\nu}+\bar{A}_{\nu}\nabla_{\mu}k^{\mu}\right)  =0. \label{eq:orderk}%
\end{equation}
To simplify the calculations further, the amplitude vector $\bar{A}^{\mu}$ can
be expressed in terms of a normalized spacelike polarization vector, $\xi
^{\mu}$, and an amplitude, $\bar{A}$, as $\bar{A}^{\mu}=\bar{A}\xi^{\mu}$.
Here, $\xi^{\mu}$ and $\bar{A}$ satisfy $\xi^{\mu}\xi_{\mu}^{\ast}=1$ and
$\bar{A}^{2}=\bar{A}^{\mu}\bar{A}_{\mu}^{\ast}$, respectively. Additionally,
expanding the Lorentz gauge with respect to $\epsilon$, the obtained leading
term implies that the polarization vector is orthogonal to the wavevector,
$k_{\mu}\xi^{\mu}=0$.

Contracting Eq. $\left(  \ref{eq:orderk}\right)  $ with $\bar{A}^{\nu\ast}$
and summing with its complex conjugate, we arrive at
\begin{equation}
\nabla_{\mu}\left[  f\left(  \phi\right)  \bar{A}^{2}k^{\mu}\right]  =0,
\label{eq:amplitude}%
\end{equation}
which is a conservation equation for the amplitude of $\bar{A}^{\mu}$ along
null geodesics. In the Einstein-Maxwell theory, where $f\left(  \phi\right)
=1$, Eq. $\left(  \ref{eq:amplitude}\right)  $ reduces to the conservation of
photon number \cite{Fleury:2015hgz}. However, due to the photon-scalar
interaction, the conservation of photon number is no longer preserved in the
EMS model. Instead, a new conserved quantity, $f\left(  \phi\right)  \bar
{A}^{2}k^{0}$, emerges. Moreover, using Eqs. $\left(  \ref{eq:orderk}\right)
$ and $\left(  \ref{eq:amplitude}\right)  $, we obtain the equation for the
polarization vector $\xi^{\mu}$,%
\begin{equation}
2f\left(  \phi\right)  k^{\mu}\nabla_{\mu}\xi_{\nu}-k_{\nu}\xi^{\mu}%
\nabla_{\mu}f\left(  \phi\right)  =0, \label{eq:pv}%
\end{equation}
which governs the evolution of $\xi^{\mu}$ along null geodesics. In the
Einstein-Maxwell theory, the second term of Eq. $\left(  \ref{eq:pv}\right)  $
vanishes, leading to the parallel transport of the polarization vector
$\xi^{\mu}$ along null geodesic. Conversely, the presence of the non-minimal
coupling between the scalar and electromagnetic fields prevents the parallel
transport of the polarization vector $\xi^{\mu}$ in the EMS model.
Consequently, a numerical method is necessary to compute the evolution of the
polarization vector $\xi^{\mu}$ along null geodesics.

\subsection{Ring Model}

To investigate polarized images of synchrotron radiation, a simplified model
of an axisymmetric, emitting ring surrounding a Schwarzschild black hole has
been proposed \cite{EventHorizonTelescope:2021btj}. Remarkably, this toy ring
model has been demonstrated to effectively capture the polarimetric image
morphology for a specific subset of GRMHD simulations of M87$^{\ast}$. In
particular, a narrow ring of radius $r_{\text{e}}$ is situated on the black
hole's equatorial plane, and radiating fluid elements within the ring emit
linearly polarized synchrotron radiation in the presence of a local magnetic
field. In the following, we briefly review the ring model, incorporating
modifications arising from the photon-scalar interaction in the EMS model. For
a comprehensive treatment of the ring model, please refer to
\cite{EventHorizonTelescope:2021btj}.

At a point $P$ on the fluid ring, a set of orthonormal tetrad $e_{\ \left(
a\right)  }^{\mu}$ can be constructed to describe the local $P$-frame,
\begin{equation}
e_{\left(  t\right)  }=\frac{e^{\delta}}{\sqrt{N}}\partial_{t},\quad
e_{\left(  r\right)  }=\sqrt{N}\partial_{r},\quad e_{\left(  \varphi\right)
}=\frac{1}{r\sin\theta}\partial_{\varphi},\quad e_{\left(  \theta\right)
}=-\frac{1}{r}\partial_{\theta}, \label{eq:tetrad}%
\end{equation}
where the negative sign ensures that $\left(  e_{\left(  r\right)
},e_{\left(  \varphi\right)  },e_{\left(  \theta\right)  }\right)  $ forms a
right-handed system. The fluid at the point $P$ moves in the $\left(
r\right)  $-$\left(  \varphi\right)  $ plane of the $P$-frame with a velocity
$\vec{\beta}$, given by
\begin{equation}
\vec{\beta}=\beta\left(  \cos\chi e_{\left(  r\right)  }+\sin\chi e_{\left(
\varphi\right)  }\right)  ,
\end{equation}
where $\chi$ is the angle between the velocity $\vec{\beta}$ and the basis
vector $e_{\left(  r\right)  }$. We assume that $\beta$ and $\chi$ are
constant on the ring. Furthermore, the fluid's local rest frame at the point
$P$, denoted as the $F$-frame, can be obtained via a boost transformation from
the $P$-frame, $e_{\ \hat{a}}^{\mu}=\Lambda_{\ \hat{a}}^{\left(  b\right)
}e_{\ \left(  b\right)  }^{\mu}$. Here, $\Lambda$ is the transformation matrix
with $\gamma=1/\sqrt{1-\beta^{2}}$,
\begin{equation}
\Lambda=\left(
\begin{array}
[c]{cccc}%
\gamma & -\beta\gamma\cos\chi & -\beta\gamma\sin\chi & 0\\
-\beta\gamma\cos\chi & \left(  \gamma-1\right)  \cos^{2}\chi+1 & \left(
\gamma-1\right)  \sin\chi\cos\chi & 0\\
-\beta\gamma\sin\chi & \left(  \gamma-1\right)  \sin\chi\cos\chi & \left(
\gamma-1\right)  \sin^{2}\chi+1 & 0\\
0 & 0 & 0 & 1
\end{array}
\right)  . \label{eq:Lorenztransformation}%
\end{equation}
In the $F$-frame, the magnetic field around the fluid has only spatial
components $\vec{B}=\left(  B^{\hat{r}},B^{\hat{\varphi}},B^{\hat{\theta}%
}\right)  $. We assume that the components of the local magnetic field in the
fluid's rest frame are constant at any point on the ring. Given the
$3$-wavevector $\vec{k}=\left(  k^{\hat{r}},k^{\hat{\varphi}},k^{\hat{\theta}%
}\right)  $ in the $F$-frame, the normalized polarization vector $\left(
\xi^{\hat{0}},\vec{\xi}\right)  $ of the synchrotron radiation emitted by the
rotating fluid can be expressed as
\begin{equation}
\xi^{\hat{0}}=0,\quad\vec{\xi}=\frac{\vec{k}\times\vec{B}}{\left\vert \vec
{k}\times\vec{B}\right\vert }. \label{eq:pvF}%
\end{equation}
Using this equation, we can determine the polarization vector $\xi_{\text{e}%
}^{\mu}$ for an emitted light ray of $4$-wavevector $k_{\text{e}}^{\mu}$.
Specifically, after acquiring the components of $k_{\text{e}}^{\mu}$ in the
$P$-frame, the boost transformation $\left(  \ref{eq:Lorenztransformation}%
\right)  $ yields its $3$-wavevector $\vec{k}_{\text{e}}$ in the $F$-frame.
Subsequently, the normalized polarization $3$-vector $\vec{\xi}_{\text{e}}$ is
determined by Eq. $\left(  \ref{eq:pvF}\right)  $. Finally, the inverse of the
boost transformation $\left(  \ref{eq:Lorenztransformation}\right)  $ and the
tetrad basis $\left(  \ref{eq:tetrad}\right)  $ provide the components of
$\xi_{\text{e}}^{\mu}$ in the spacetime coordinates.

In the case of an optically thin medium, the emitted intensity in the
$F$-frame depends on the angle $\zeta$ between the fluid's $3$-wavevector
$\vec{k}_{\text{e}}$ and the magnetic field $\vec{B}$, the frequency $\nu$ of
the emitted photons and the geodesic path length $l_{p}$ in the emitting
region. Specifically, the emitted intensity is given by
\begin{equation}
I_{\nu,\text{e}}\propto\nu^{-\alpha_{\nu}}l_{p}\left\vert \vec{B}\right\vert
^{1+\alpha_{\nu}}\sin^{1+\alpha_{\nu}}\zeta.
\end{equation}
In models of M87$^{\ast}$, a dependence of $\sin^{2}\zeta$ is often observed
at $230$ GHz. Consequently, we set $\alpha_{\nu}=1$ for the remainder of this
paper. In the Einstein-Maxwell theory, the conservation of photon number can
be used to calculate the intensity $I_{\nu}$ along geodesics. Conversely, in
the EMS model, the aforementioned quantity $f\left(  \phi\right)  \bar{A}%
^{2}k^{0}$ is conserved, resulting in a new invariant along the geodesics,
$f\left(  \phi\right)  I_{\nu}/\nu^{3}$.

If an observer at $r=r_{\text{o}}$ receives a light ray with the
$4$-wavevector $k_{\text{o}}^{\mu}$, we can use the backward ray-tracing
method to determine the 4-wavevector $k_{\text{e}}^{\mu}$ of the light ray
when it is emitted from the ring. Specifically, the light ray is obtained by
numerically integrating the geodesic equations,
\begin{align}
\frac{dx^{\mu}}{d\lambda} &  =k^{\mu},\nonumber\\
\text{ }\frac{dk^{\mu}}{d\lambda} &  =-\Gamma_{\rho\sigma}^{\mu}k^{\rho
}k^{\sigma},
\end{align}
where $\lambda$ represents the affine parameter, and $\Gamma_{\rho\sigma}%
^{\mu}$ is the Christoffel symbol. Once $k_{\text{e}}^{\mu}$ is given, the
emitted polarization vector $\xi_{\text{e}}^{\mu}$ can be determined as
previously discussed. Eq. $\left(  \ref{eq:pv}\right)  $ is then used to
evolve $\xi^{\mu}$ forward along the light ray, yielding the observed
polarization vector $\xi_{\text{o}}^{\mu}$. Additionally, the observed
intensity $I_{\nu,\text{o}}$ is derived from the invariant $f\left(
\phi\right)  I_{\nu}/\nu^{3}$,
\begin{equation}
I_{\nu,\text{o}}=\frac{f\left[  \phi\left(  r_{\text{e}}\right)  \right]
}{f\left[  \phi\left(  r_{\text{o}}\right)  \right]  }\delta^{3}%
I_{\nu,\text{e}}\propto\frac{f\left[  \phi\left(  r_{\text{e}}\right)
\right]  }{f\left[  \phi\left(  r_{\text{o}}\right)  \right]  }\frac
{\delta^{3}}{k_{\text{e}}^{\hat{t}}}l_{p}\left\vert \vec{B}\right\vert
^{2}\sin^{2}\zeta,\label{eq:intensity}%
\end{equation}
where $\delta=\nu_{\text{o}}/k_{\text{e}}^{\hat{t}}$ is the Doppler factor,
and $\nu_{\text{o}}$ is the observed frequency.

Furthermore, we assume that the observer is located in a frame with the
observer basis $\left(  e_{\left(  t\right)  },e_{\left(  r\right)
},e_{\left(  \theta\right)  },e_{\left(  \phi\right)  }\right)  $. In this
observer's frame, the direction of the measured polarization vector is
indicated by the EVPA, defined as,
\begin{equation}
\text{EVPA}=\arctan\left(  -\frac{\xi_{\text{o}}^{\left(  \varphi\right)  }%
}{\xi_{\text{o}}^{\left(  \theta\right)  }}\right)  ,
\end{equation}
where $\xi_{\text{o}}^{\left(  a\right)  }$ represents the components of the
polarization vector in the observer's frame. Additionally, the observed
frequency $\nu_{\text{o}}$ corresponds to the $\left(  t\right)  $-component
of the observed wavevector, $k_{\text{o}}^{\left(  t\right)  }$, which can be
conveniently set to $1$ without compromising generality. The observation
angles $\Theta$ and $\Phi$ of received light rays are defined as
\cite{Cunha:2016bpi},
\begin{equation}
\sin\Theta=\frac{k_{\text{\text{o}}}^{\left(  \theta\right)  }}{k_{\text{o}%
}^{\left(  t\right)  }},\quad\tan\Phi=\frac{k_{\text{o}}^{\left(  \phi\right)
}}{k_{\text{o}}^{\left(  r\right)  }}.
\end{equation}
Consequently, the Cartesian and polar coordinates in the image plane are
expressed as
\begin{equation}
x=-r_{\text{o}}\Phi,\quad y=r_{\text{o}}\Theta,
\end{equation}
and
\begin{equation}
\rho=\sqrt{x^{2}+y^{2}},\quad\varphi_{\rho}=\arccos\frac{x}{\rho},
\end{equation}
respectively. In the image plane, the EVPA is measured counterclockwise
relative to the positive $y$-axis.

\section{Results}

\label{sec:Numerical-Result}

In this section, we investigate polarized images of the aforementioned
emitting ring around scalarized RN black holes. Our numerical simulations
model the ring as a torus with a minor radius of $H=0.01$ and a major radius
equal to the ISCO radius, $r_{\text{ISCO}}$, as detailed in
\cite{Chen:2024ilc}. The torus's center lies on the equatorial plane. For
simplicity, we approximate the geodesic path length as $l_{p}=Hk^{\hat{t}%
}/k^{\hat{z}}$. The observer is positioned at a radius of $r_{\text{o}}=100$
with an inclination angle of $\theta_{\text{o}}=20^{\circ}$, mimicking the
observational setup for M87$^{\ast}$.

To simulate observational images, we vary the observer's viewing angles,
$\Theta$ and $\Phi$, and numerically integrated at least $2000\times2000$
photon trajectories until they intersected the narrow ring or reached the
cutoff radii at $r=r_{+}+0.001$ or $110$. In this paper, we focus exclusively
on the primary image of the ring, deferring the analysis of higher-order
images to future studies. Once the light rays are determined, the polarization
vector and intensity are calculated using Eqs. $\left(  \ref{eq:pv}\right)  $
and $\left(  \ref{eq:intensity}\right)  $, respectively. To reduce numerical
fluctuations arising from the finite grid points, we employed the
$MeanShiftFilter$ function of $Wolfram\text{ }Mathematica$.

\subsection{Equatorial Magnetic Field}

The EHT Collaboration's polarized image of M87$^{\ast}$ indicates that, when
the black hole's tilt is oriented northward, the polarized flux is
concentrated in the right half of the image plane \cite{Akiyama:2021qum}.
Furthermore, the twist angle between the EVPA and the local radial unit vector
remains relatively stable in regions of high polarized flux. Previous studies
have shown that an equatorial local magnetic field with non-zero $B^{\hat{r}}$
and $B^{\hat{\phi}}$ components can effectively reproduce these key features
\cite{EventHorizonTelescope:2021btj}. Inspired by these findings, we begin our
investigation by analyzing simulations of the ring model, where $B^{\hat{r}}$
and $B^{\hat{\phi}}$ are non-zero, while $B^{\hat{z}}$ is set to zero. Given
the probable parallel alignment of the equatorial magnetic field and the fluid
motion, we assume that the fluid's velocity is aligned with the magnetic
field. Specifically, we define the angle $\chi$ as%
\begin{equation}
\chi=\arctan\frac{B^{\hat{\varphi}}}{B^{\hat{r}}}+\pi.
\end{equation}
It is worth noting that for an equatorial magnetic field, the alternative
choice, $\chi=\arctan\left(  B^{\hat{\varphi}}/B^{\hat{r}}\right)  $, results
in the same linear polarized emission.

\begin{figure}[ptb]
\begin{centering}
\includegraphics[width=1\textwidth]{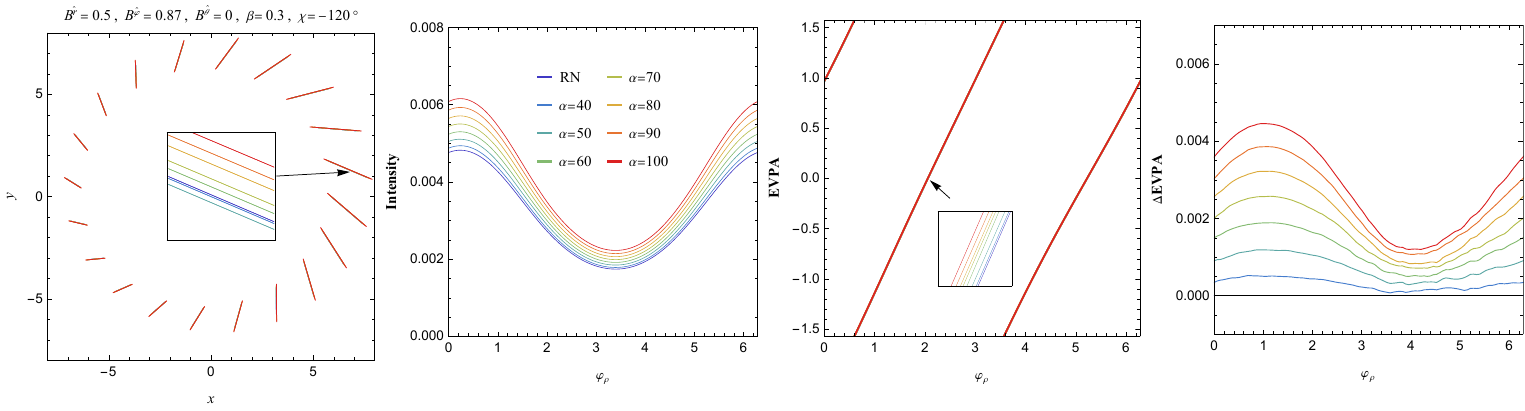}
\par\end{centering}
\begin{centering}
\includegraphics[width=1\textwidth]{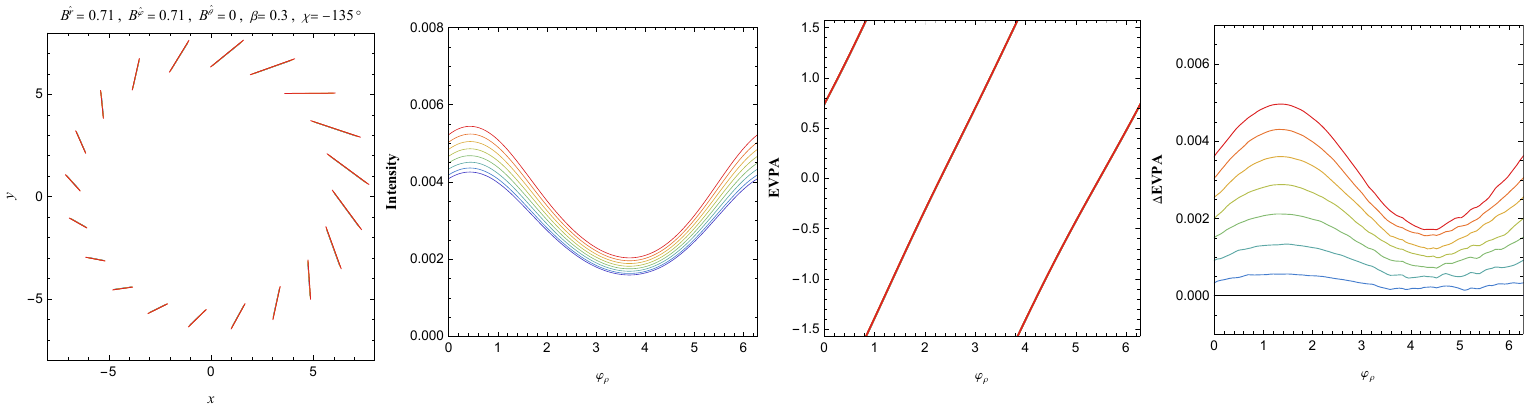}
\par\end{centering}
\centering{}\includegraphics[width=1\textwidth]{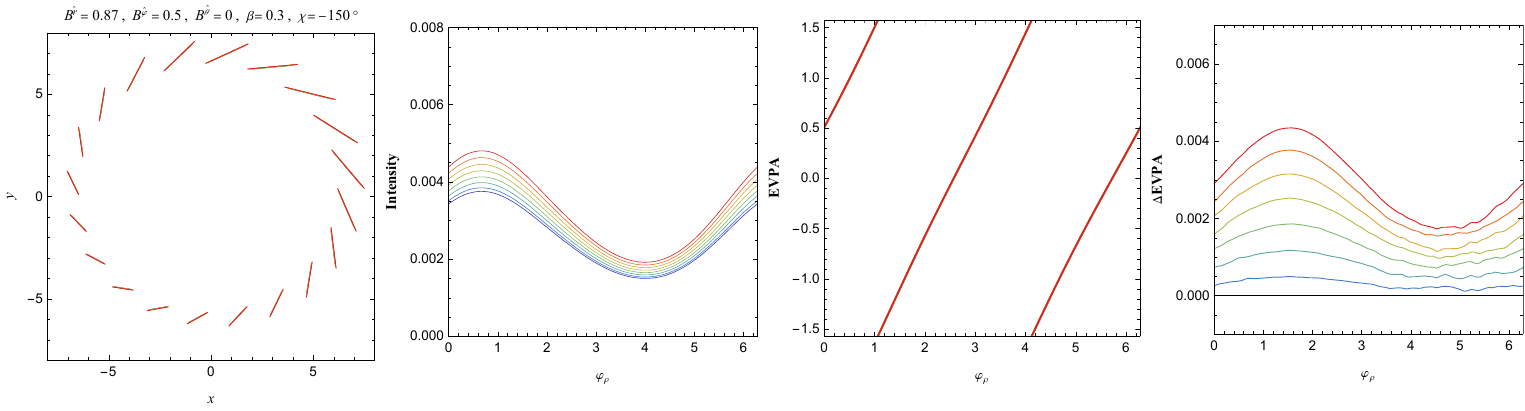}
\caption{Linear polarization of a synchrotron-emitting ring surrounding RN and
scalarized RN black holes for a fixed charge, $Q=0.4$, in three configurations
of equatorial magnetic fields. The configurations correspond to\ $\chi
=-120^{\circ}$ (\textbf{Top Row}), $-135^{\circ}$ (\textbf{Middle Row}) and
$-150^{\circ}$ (\textbf{Bottom Row}), with the magnetic field oriented
opposite to the fluid's velocity. For scalarized RN black holes, the coupling
constant $\alpha$ assumes the values of $40$, $50$, $60$, $70$, $80$, $90$ and
$100$. The first column presents the polarized images, which exhibit flux
asymmetry qualitatively consistent with observations of M87$^{\ast}$. The
second, third and fourth columns display the polarization intensity, the
polarization direction EVPA and the deviation of the EVPA from RN black holes,
$\Delta$EVPA, respectively, as functions of the azimuthal angle $\varphi
_{\rho}$. As $\alpha$ increases, both the intensity and the EVPA of scalarized
RN black holes rise.}%
\label{Fig:Eq04}%
\end{figure}

Fig. $\ref{Fig:Eq04}$ presents simulations of the observable polarization for
RN and scalarized RN black holes, considering three distinct configurations of
purely equatorial magnetic fields. These configurations correspond to fluids
flowing inward in a clockwise direction at angles of $\chi=-120^{\circ}$,
$-135^{\circ}$ and $-150^{\circ}$, as illustrated in the top, middle and
bottom rows, respectively. The left column displays the polarized images,
where polarization vectors are represented by $20$ equally spaced ticks along
the azimuthal angle $\varphi_{\rho}$ on the ring image. The length of the
polarization vectors signifies the intensity, while their twist around the
ring encodes the EVPA. These images are consistent with observations of
M87$^{\ast}$, exhibiting the most intense flux concentrated on the right-hand
side. Notably, the apparent size of the ring images exhibits a weak dependence
on the coupling constant $\alpha$, resulting in minimal discernible
differences in the polarization vectors at a given $\varphi_{\rho}$. To
highlight these differences, the upper-left panel includes a zoomed-in inset,
revealing that a stronger $\alpha$ can yield a larger apparent radius.

The second column from the left presents the polarization intensity as a
function of $\varphi_{\rho}$ for RN and scalarized RN black holes. Although
the intensity maintains a largely similar distribution with peak values in the
top right quadrant ($0<\varphi_{\rho}<\pi/2$), its magnitude increases with
rising coupling constant $\alpha$ or tangential magnetic field component
$B^{\hat{\phi}}$. The third and fourth columns, respectively, illustrate the
polarization direction EVPA and the relative direction angle, $\Delta
$EVPA$\equiv$EVPA$-$EVPA$_{\text{RN}}$, as functions of $\varphi_{\rho}$. For
scalarized RN black holes, the EVPA deviations from RN black holes are
relatively minor. Furthermore, within a given magnetic field configuration,
the EVPA deviations increase with growing $\alpha$, causing the polarization
vector to rotate slightly towards the tangential direction. As the radial
magnetic field component diminishes, the EVPA at a fixed $\varphi_{\rho}$
decreases, indicating a growing tangential component of the polarization vectors.

\begin{figure}[ptb]
\begin{centering}
\includegraphics[width=1\textwidth]{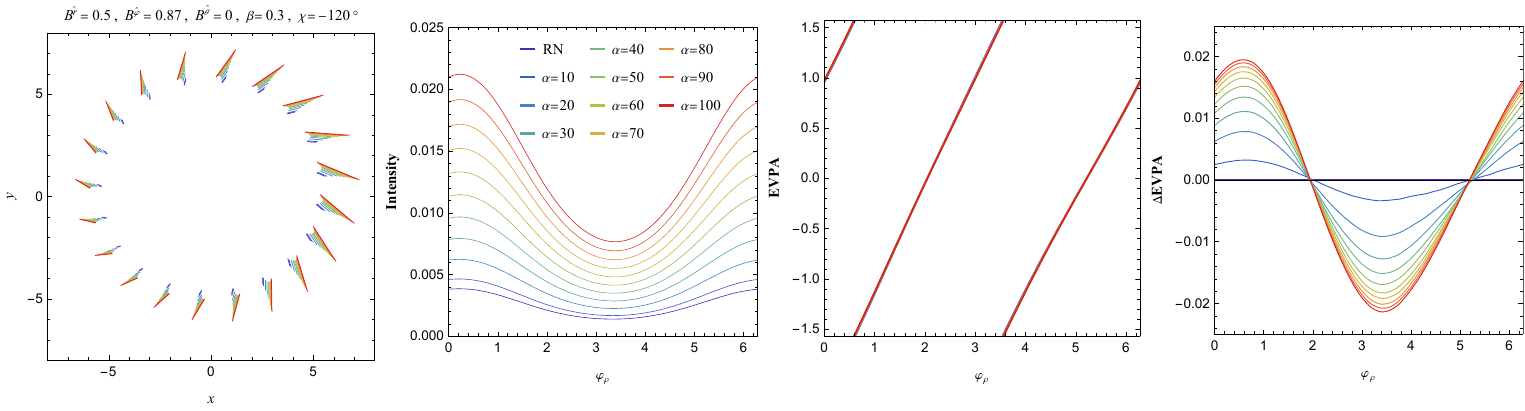}
\par\end{centering}
\begin{centering}
\includegraphics[width=1\textwidth]{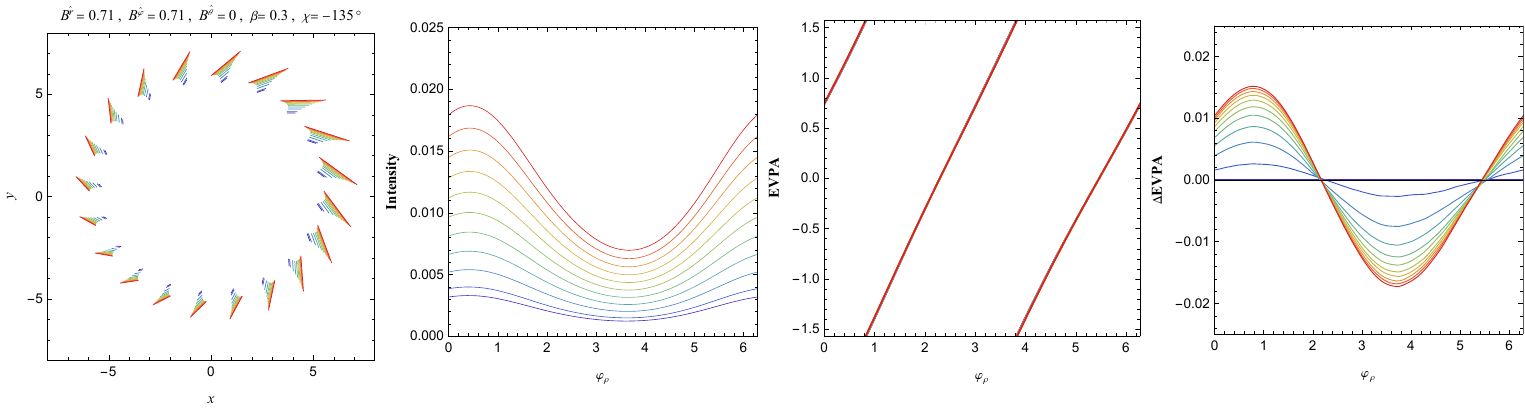}
\par\end{centering}
\centering{}\includegraphics[width=1\textwidth]{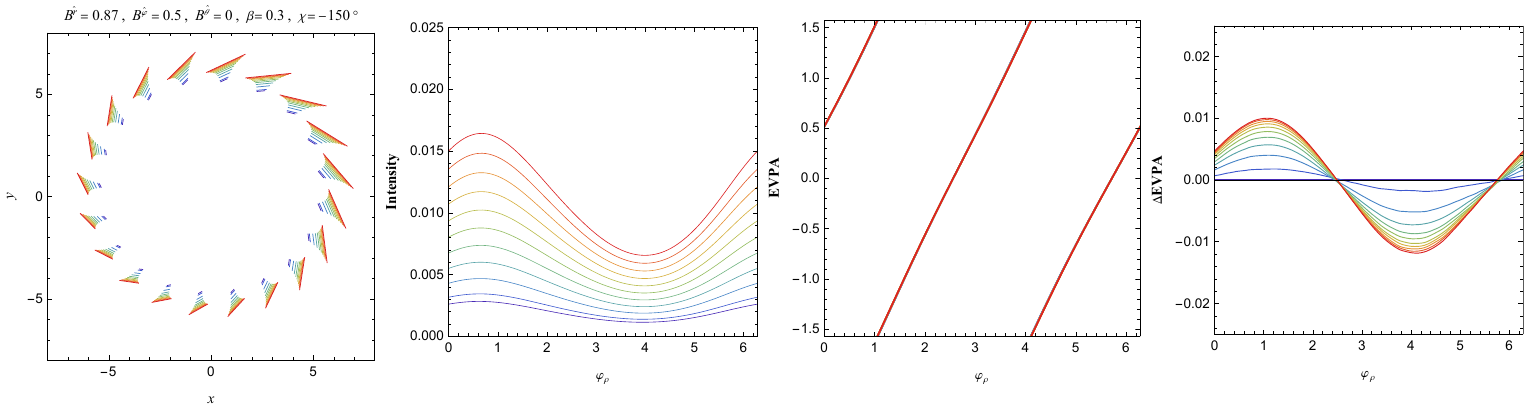}\caption{Linear
polarization of a synchrotron-emitting ring surrounding RN and scalarized RN
black holes for $Q=0.9$, considering the same equatorial magnetic field
configurations as described previously. The pattern of polarized images
closely resembles the $Q=0.4$ case. Compared to the $Q=0.4$ case, the
intensity flux, represented by the length of polarization vectors, increases
more substantially with growing $\alpha$.}%
\label{Fig:Eq09}%
\end{figure}

Fig. $\ref{Fig:Eq09}$ depicts the polarized images, intensity and EVPA for RN
and scalarized RN black holes with $Q=0.9$, considering the same magnetic
field configurations as in Fig. $\ref{Fig:Eq04}$. Similar to the $Q=0.4$ case,
the polarized images exhibit flux asymmetry, with the most intense polarized
flux concentrated in the upper-right region. However, a larger black hole
charge $Q$ introduces several notable differences. Firstly, the apparent
radius of the ring images becomes more sensitive to $\alpha$, making it easier
to distinguish the polarization vectors at a fixed $\varphi_{\rho}$ for
different $\alpha$ values. The ring image with a stronger $\alpha$ exhibits a
larger apparent radius. Secondly, due to the smaller ISCO radius, the
polarization intensity of scalarized RN black holes is significantly
amplified, leading to more pronounced deviations from RN black holes. For
instance, the intensity with $\alpha=100$ can be up to four times that of the
RN black hole. Finally, the discrepancy in the EVPA, $\Delta$EVPA, of
scalarized RN black holes also becomes more pronounced. In regions of weak
intensity, the EVPA decreases as $\alpha$ increases, rather than increasing.

\subsection{Vertical Magnetic Field}

\begin{figure}[ptb]
\begin{centering}
\includegraphics[width=1\textwidth]{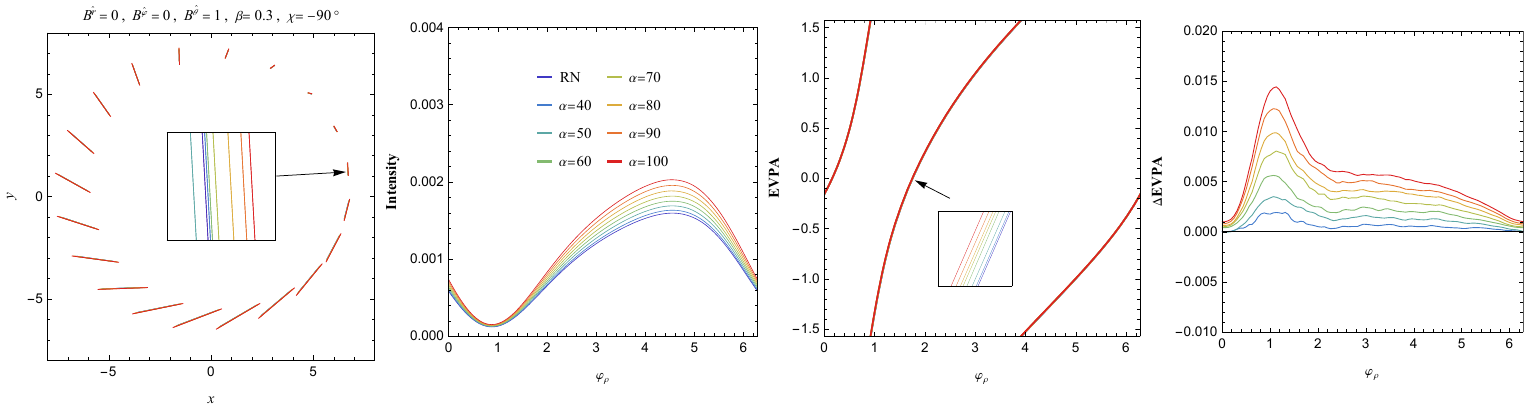}
\par\end{centering}
\begin{centering}
\includegraphics[width=1\textwidth]{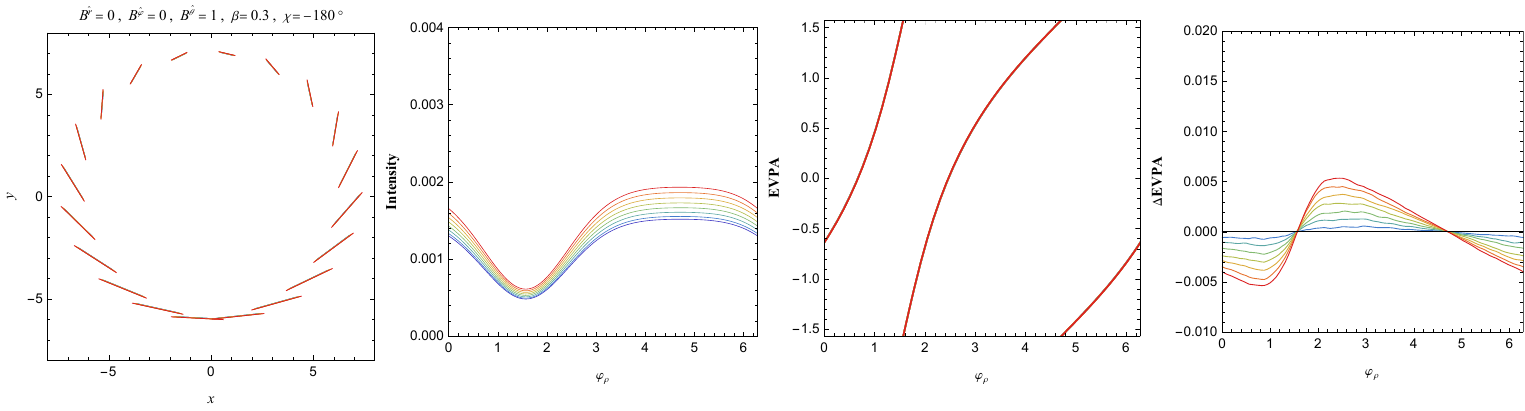}
\par\end{centering}
\centering{}\caption{Linear polarization of a synchrotron-emitting ring
surrounding RN and scalarized RN black holes for $Q=0.4$ in a vertical
magnetic field. The upper and lower rows correspond to purely clockwise
rotation ($\chi=-90^{\circ}$) and purely radial infall ($\chi=-180^{\circ}$),
respectively. The most prominent intensity flux is concentrated at the bottom
of the polarized images.}%
\label{fig:Vq04}%
\end{figure}

\begin{figure}[ptb]
\begin{centering}
\includegraphics[width=1\textwidth]{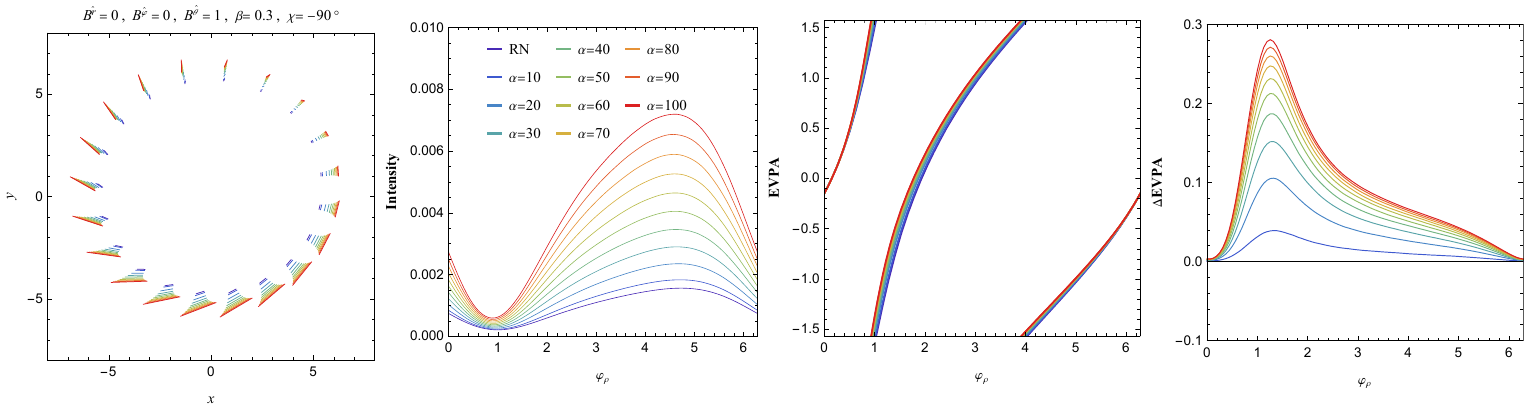}
\par\end{centering}
\begin{centering}
\includegraphics[width=1\textwidth]{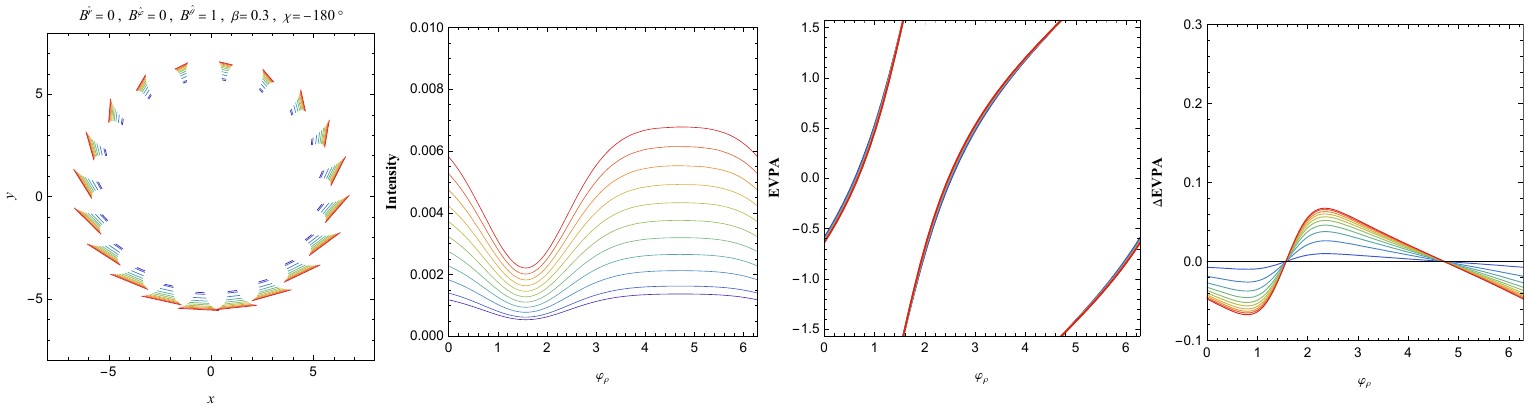}
\par\end{centering}
\centering{}\caption{Linear polarization of a synchrotron-emitting ring
surrounding RN and scalarized RN black holes for $Q=0.9$ in a vertical
magnetic field. The upper and lower rows correspond to purely clockwise
rotation ($\chi=-90^{\circ}$) and purely radial infall ($\chi=-180^{\circ}$),
respectively. Although the polarization pattern closely resembles the $Q=0.4$
case, the overall intensity and $\Delta$EVPA are significantly greater.}%
\label{fig:Vq09}%
\end{figure}

As demonstrated in \cite{EventHorizonTelescope:2021btj}, the polarization
pattern of the ring model in the Schwarzschild spacetime with a vertical
magnetic field is inconsistent with the observations of M87$^{\ast}$,
primarily due to aberration effects. To explore whether this pattern persists
for scalarized RN black holes, we analyze cases with a purely vertical
magnetic field for scalarized RN black holes with $Q=0.4$ and $0.9$, as
presented in Figs. $\ref{fig:Vq04}$ and $\ref{fig:Vq09}$, respectively. The
upper and lower rows in these figures correspond to fluid flows that are
clockwise tangential ($\chi=-90^{\circ}$) and radially inward ($\chi
=-180^{\circ}$), respectively.

The left column of Figs. $\ref{fig:Vq04}$ and $\ref{fig:Vq09}$ presents the
corresponding polarized images, with ticks representing polarization vectors.
Consistent with the behavior of Schwarzschild black holes, our numerical
simulations demonstrate that the regions of strongest intensity flux in these
images are concentrated at the bottom, with the intensity diminishing upward.
This observation is further supported by the polarization intensity as a
function of $\varphi_{\rho}$, as depicted in the second column. Similar to the
equatorial magnetic field case, the intensity of scalarized RN black holes and
their deviations from RN black holes are more pronounced for larger $Q$
values. However, the overall intensity of the vertical magnetic field is
significantly lower than that of the equatorial magnetic field. The third and
fourth columns present the EVPA and $\Delta$EVPA as functions of
$\varphi_{\rho}$ for various $\alpha$ values. While the EVPA exhibits minimal
sensitivity to $\alpha$ for $Q=0.4$, its dependence on $\alpha$ becomes more
apparent when $Q=0.9$. Furthermore, as $\alpha$ increases, the EVPA
consistently increases in the purely clockwise rotation configuration. In
contrast, for the purely radial infall configuration, as $\alpha$ grows, the
EVPA increases in regions where the intensity rises with azimuthal angle
$\varphi_{\rho}$ but decreases in regions where the intensity falls with
azimuthal angle $\varphi_{\rho}$.

\section{Conclusions}

\label{sec:Conclusions}

In this study, we have examined the linear polarization of
synchrotron-emitting rings surrounding both RN and scalarized RN black holes,
focusing on their observable characteristics in the presence of equatorial or
vertical magnetic fields. Our analysis reveals that the polarization patterns
of RN and scalarized RN black holes share qualitative similarities with those
of Schwarzschild black holes. Specifically, we observed flux asymmetry in the
polarized images, with the most intense flux region typically concentrated in
the upper-right and bottom regions for equatorial and vertical magnetic
fields, respectively.

Furthermore, we investigated the effects of varying black hole charge $Q$,
coupling constant $\alpha$ and magnetic field configurations on the
polarization patterns of scalarized RN black holes. For larger $Q$ values, we
observed a significant increase in intensity flux and deviations of the EVPA,
further differentiating scalarized RN black holes from their RN counterparts.
As $\alpha$ increases, the polarized images exhibit notable changes, including
larger apparent radii, higher polarization intensity and greater EVPA
deviations. Additionally, equatorial magnetic fields generally yield higher
intensity and distinct EVPA variations compared to vertical magnetic fields.

In conclusion, this research contributes to our comprehension of polarization
signatures of black holes in the presence of scalar fields and magnetic
fields. The insights derived from these simulations may serve as valuable
benchmarks for interpreting observational data from existing and future
astronomical instruments. Future research endeavors could focus on
higher-order images, particularly in the context of double photon spheres.
Moreover, scalar clouds surrounding Kerr-Newman black holes and scalarized
Kerr-Newman black holes have been recently constructed
\cite{Guo:2023mda,Guo:2024bkw}. Consequently, extending our analysis presented
in this paper to the rotating case is highly desirable.

\begin{acknowledgments}
We are grateful to Tianshu Wu for useful discussions and valuable comments.
This work is supported in part by NSFC (Grant Nos. 12105191, 12275183,
12275184 and 11875196).
\end{acknowledgments}

\bibliographystyle{unsrturl}
\bibliography{ref}

\end{document}